\documentstyle[prl,aps,epsf,twocolumn]{revtex}

\title
{
de-Broglie Wave-Front Engineering
}

\author
{
M. Olshanii, N. Dekker, C. Herzog, and M. Prentiss
}
\address
{
Lyman Laboratory, Harvard University,
Cambridge, MA 02138, USA\\
{\small E-mail: {\it maxim@atomsun.harvard.edu}}
}

\date{\today}

\begin{document}
\maketitle
\begin{abstract}
We propose a simple method  for the deterministic generation of an
{\it arbitrary} continuous quantum state 
of the center-of-mass of an atom.
The method's spatial resolution gradually increases with 
the interaction time with no apparent fundamental limitations.
Such {\it de-Broglie Wave-Front Engineering} of the 
atomic density can find applications in Atom Lithography,
and we discuss possible implementations of our scheme in  
atomic beam experiments. 
\end{abstract}
%
%




Engineering of quantum states has been a widely discussed topic 
for the last decade. Apart from  a purely academic interest,
there exist numerous applications of quantum state engineering 
including preparation of non-classical states of a  cavity
electromagnetic field, ``programming'' of a trapped-ion-based
quantum computer, and atom lithography. Initial theoretical 
suggestions \cite{Kurizki_1,Akulin_1,Kurizki_2,Kurizki_3}
for the preparation of a {\it pre-chosen} quantum state
of a cavity field were based on the so-called conditional measurement
method, where the target state is reached after a ``successful'' sequence 
of quantum measurements, while the ``unsuccessful'' measurement events
are discarded. In the schemes 
\cite{Law_1,Cirac_1,Law_3}, applicable to both 
cavity light and external motion of a trapped ion, a two-level
atom, coupled to the quantum field of interest as well as to a controllable 
external laser light, plays a role of a ``bus'', which  transfers, 
in a prescribed way,
population and coherence between the discrete eigenstates 
of the quantum field. Similar ideas 
were used to generate an arbitrary internal state of a multilevel atom
\cite{Law_2}.
According to the suggestions \cite{Marte_1,Marte_2}, the adiabatic
population transfer process allows a one-to-one mapping 
between a quantum state of a Zeeman multiplet and 
a cavity field.

The quantum state engineering methods listed above  
deal with systems of a discrete spectrum. In our paper we suggest 
a simple method to create an arbitrary 
{\it continuous} motional state of a free atom starting from a plane wave
as an initial condition. The role of a bus, transferring the coherence 
between the initial and ``target'' states, is played by an
external uniform force field: the ``target'' motional state
is encoded in the time dependence of the amplitude of  
applied laser light.


The general idea of  the set-up for the realization of our
{\it de-Broglie Wave-Front Engineering} scheme was inspired 
by the precision position measurement technique suggested and experimentally 
realized by 
J. E. Thomas \cite{Thomas}. Let us consider a two-level atom, interacting 
with a magnetic field $H(z) = -\alpha z$ whose amplitude 
varies linearly in space. Suppose that the internal
atomic state $|1\rangle$ does not interact with the magnetic field
(the corresponding Lande factor equals zero: $g_{1} = 0$), 
whereas the Lande factor for the state $|2\rangle$ has a 
finite value $g_{2} = g$. The energy difference between 
the states $|2\rangle$ and $|1\rangle$ will, thus, depend linearly on the 
position of the atom:
\begin{eqnarray} 
\hbar \omega_{2,1}(z) = \hbar \omega_{2,1}(0) - {\cal F}z \, ,
\end{eqnarray}
where $\hbar \omega_{2,1}(0) = E_{2} - E_{1}$ is the 
energy difference between $|2\rangle$ and $|1\rangle$ 
in the absence of the magnetic 
field, and the gradient force acting on atom in state $|2\rangle$ is
${\cal F} = \alpha \mu_{\rm Bohr} g$. In what follows 
we will assume, without loss of generality, 
that both $\alpha$ and $g$ are positive numbers. 

\begin{figure}
\begin{center}
\leavevmode
\epsfxsize=0.45\textwidth
\epsffile{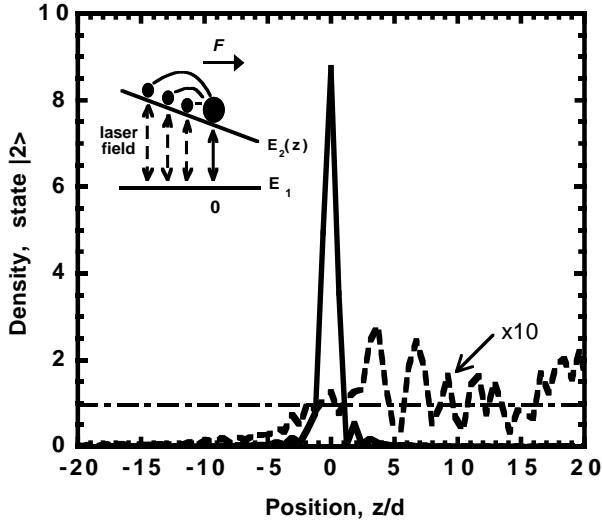}
\end{center}
\caption
{
Spatial distribution of atoms in the state $|2\rangle$ 
({\it ab initio}  quantum-mechanical simulation). The desired pattern
corresponds to a $\delta$-functional peak centered at $z=0$.
Dashed line is a result of an application of a monochromatic
field at resonance at $z=0$.
Solid line shows the ``de-Broglie wave-front engineering'' result,
where the frequency was chirped according to 
``classical'' {\it ansatz} (\ref{V(t)},\ref{trajectory})
taken at $p_{0} = 0$.
Dashed-dotted line shows the initial distribution of the 
state $|1\rangle$ atoms.
Interaction time is $T = 4.8 \times \tau$, field amplitude  is
$\tilde{V} = 1. \times (\hbar/\tau)$. The characteristic length $d$ and 
time $\tau$ of the problem are defined by the expressions
(\ref{gravitational_width}) and (\ref{gravitational_time}) respectively.
The inset illustrates the basic idea of the 
``de-Broglie Wave-Front Engineering'' method: the trajectory
of the resonant point is designed in such a way that state $|2\rangle$
atoms created at different stages of the process come to the target
at the same time $t=T$.
\label{f_single}
}
\end{figure}

Suppose for a moment that our goal is to create 
the narrowest possible position distribution of atoms in  
the state $| 2 \rangle$ centered at a position $z = z^{\star}$,
and that the initial condition   
corresponds to a state $|1\rangle$ atom of some 
momentum $p_{0}$ (whose value we can adjust at will).
The simplest (but as we will see below not the optimal) way
to approach the above goal is to apply a monochromatic 
spatially uniform laser field of a frequency 
$\omega = \omega_{2,1}(z^{\star})$
for some period of time $T$ (solid vertical arrow, inset for the 
Fig.~\ref{f_single}). 
As it is shown in the work \cite{Thomas},
for a given value of the force ${\cal F}$ the spatial width 
of the distribution of atoms created in the state $|2\rangle$
is limited from below by a value
\begin{eqnarray}
d = \left(\frac{\hbar^2}{2M{\cal F}}\right)^{\frac{1}{3}}
\label{gravitational_width}
\end{eqnarray}
(the so-called {\it diffraction limit}) no matter how long 
the interaction time $T$ is. This limit is reached at 
a time of the order of 
\begin{eqnarray}
\tau = \left(\frac{2 \hbar M}{{\cal F}^2}\right)^{\frac{1}{3}}
\label{gravitational_time}
\end{eqnarray}
At times shorter than the time (\ref{gravitational_time}),
a wave packet of a minimum position-momentum uncertainty
relation ($\delta z \delta p \sim \hbar$)
is created. The external force ${\cal F}$ broadens the 
momentum distribution according to  $\delta p \sim {\cal F}T$, 
and, therefore,  the spatial width of the 
state $|2\rangle$ distribution decreases with time 
as $\delta z \sim \hbar/{\cal F}T$. 
For long interaction times 
though, the peak width starts 
increasing quadratically as the interaction time increases
(see for example the spatial
distribution calculated for the case $T = 4.8 \tau$ shown at the
Fig.~\ref{f_single}). 
Such a broadening is caused by both the quantum-mechanical
diffraction of the wave-packet being prepared and the acceleration 
of the wave-packet. In what follows we will show that it is possible to
suppress both diffraction and acceleration broadening  using 
a simple modification of the time dependence of the laser
field amplitude 
(which was 
time-independent in the above example).  

Let us write the field amplitude $V(t)$ in a form
\begin{eqnarray}
V(t) = \tilde{V} \,
e^
{
  -i\int_{0}^{t} 
  \omega_{2,1}(z_{s}(t^{\prime}, p_{0}, z^{\star})) \, dt^{\prime}  
}  ,
\label{V(t)}
\end{eqnarray}
where 
the trajectory of the ``resonant'' point
$z_{s}(t, p_{0}, z^{\star})$ should be optimized 
in such a way that at the final time $T$ the state $|2\rangle$ atoms
will form a narrow peak, centered at $z=z^{\star}$.  
Notice that at a given time $t$ the field (\ref{V(t)})
plays a role of a spatially localized source of atoms in internal
state $|2\rangle$ and with momentum $p_0$.
Consider then a classical analog of our problem: 
\begin{description}
\item
\noindent
Find a trajectory $z_{s}(t, p_{0}, z^{\star})$
of a classical source of atoms of an initial momentum $p_{0}$, 
such that all the atoms 
emitted will reach the ``target'' $z=z^{\star}$ at a 
preselected time $T$. Atoms are supposed to be affected by a
force ${\cal F}$. 
\end{description}
Such a trajectory does exist: it is given by
\begin{eqnarray}
z_{s}(t, p_{0}, z^{\star}) 
= 
z^{\star} - \frac{p_{0}(T-t)}{M} - \frac{F(T-t)^2}{2M} \, .
\label{trajectory}
\end{eqnarray}

Let us now insert the 
{\it ansatz} (\ref{trajectory}) to the expression for the field 
amplitude (\ref{V(t)}) and evaluate the equations of motion
using this amplitude. 
At time $T$ the state $|2\rangle$ component of the 
atomic wave function will be given by  
\begin{eqnarray}
&&\lbrack \psi_{2}(z) \rbrack_{t=T}
= 
\frac{\sqrt{\rho_{\rm in}}\tilde{V}}{\hbar{\cal F}}
\int_{p_{0}}^{p_{0}+{\cal F}T}
\!\! dp \, e^{ip(z-z^{\star})/\hbar}
\label{psi_chirp}\\
&&\hspace{.5cm}
= 
          \left(
            \frac{\tilde{V}T}{\hbar}
          \right)
          \sqrt{\rho_{\rm in}} \,
          e^
          {i
          \lbrack
            (p_{0}+{\cal F}T/2)z/\hbar 
          \rbrack
          } \,
          {\rm sinc}\!\!
          \left\lbrack
                \frac{z-z^{\star}}{\delta z(T)}
          \right\rbrack \, ,
\nonumber
\end{eqnarray}
where for {\it all} interaction times $T$, the spatial width of the 
distribution reads
$
\delta z(T) = 2\hbar/FT 
$,
and {\it no diffraction or acceleration broadening is present}
(solid line at Fig. \ref{f_single}). 
Here  $\rho_{\rm in}$ is the initial density of atoms in the state $|1\rangle$.
(Above we have omitted some insignificant constant phase factors depending
on the definition of the internal states $|1\rangle$ and $|2\rangle$.)
Note that the {\it sinc}-shape is ultimately the 
best approximation for a $\delta$-functional peak
$\delta(z-z^{\star})$ one can create  
using a $\lbrack p_{0}; \, p_{0}+{\cal F}T \rbrack$ window in
momentum space.

The {\it ansatz} (\ref{V(t)},\ref{trajectory}) motivates 
our strategy for ``de-Broglie wave-front engineering'' of 
motional quantum states. Imagine that one's goal is to 
prepare an atom in a motional state $\phi(z)$ (normalized to unity:
$\int dz \, |\phi|^2 = 1$). 
Let us represent the ``target'' state as a continuous superposition of the 
$\delta$-peaks: 
$
\phi(z) = \int_{-\infty}^{+\infty} \! 
dz^{\star} \, \delta(z-dz^{\star}) \phi(z^{\star})
$.
The state engineering process 
will involve then the following steps:
\begin{enumerate}
\item
Prepare the atom in the internal state $|1\rangle$ and 
in an external state corresponding to the $p=p_{0}$ eigen-state 
of the atomic momentum
\begin{eqnarray}
\lbrack \psi_{1}(z)\rbrack _{t=0} = 
\sqrt{\rho_{\rm in}} e^{+ip_{0}z/\hbar} \, ,
\end{eqnarray}
where the initial momentum $p_{0}$ is chosen in such a way 
that the momentum window $\lbrack p_{0}; \, p_{0}+{\cal F}T \rbrack$ 
covers entirely the 
momentum distribution 
\begin{eqnarray}
\bar{\phi}(p) = \frac{1}{2\pi\hbar}
\int_{-\infty}^{+\infty} dz \, e^{-ipz/\hbar} \phi(z)
\end{eqnarray}
of the ``target'' state, ${\cal F}$ and $T$ being the typical magnetic field 
gradient and typical interaction time available in the given implementation;
\item
Apply for a time $T$ a laser field 
\begin{eqnarray}
V(t) = \tilde{V}(t)
e^
{
  -i\int_{0}^{t} 
  \omega_{2,1}(z_{s}(t^{\prime}, p_{0}, 0)) \, dt^{\prime}
} \, ,
\label{V(t)_2}
\end{eqnarray}
where 
\begin{eqnarray}
\tilde{V}(t) 
&&= \frac{
           \tilde{V}_{0} {\cal F} T
         }
         {
            2\pi\hbar\phi(0)
         } 
\int_{-\infty}^{+\infty} dz^{\star} \,
e^
{
  -ip(T,p_{0}) z^{\star}
} \times 
\nonumber\\
&& \hspace{1cm}
e^
{
  -i\int_{0}^{t}
  \lbrack
      \omega_{2,1}(z_{s}(t, p_{0}, z^{\star}))
      -
      \omega_{2,1}(z_{s}(t, p_{0}, 0)) 
  \rbrack
  \, dt^{\prime}
} 
\nonumber\\
&&=
\frac
{
  \tilde{V}_{0} {\cal F} T
}
{
  \phi(0))
} 
\times \,
\bar{\phi}(p(t,p_{0})) 
\label{V(t)_22}
\end{eqnarray}
is the field amplitude adjusted to the ``target'' state,
and $p(t,p_{0}) = p_{0} + {\cal F}(T-t)$;
\item
At the time $T$ measure the internal state.
If the atom remains in the state $|1\rangle$, repeat 
the above steps. If atom is detected in the state $|2\rangle$,
the preparation procedure is complete. 
\end{enumerate}
After a lengthy but straightforward calculation 
one can show that in the course of the preparation 
procedure the initial state gets transformed to a state 
\begin{eqnarray}
&&\lbrack \psi_{2}(z) \rbrack _{t=T} = 
\int_{p_{0}}^{p_{0}+{\cal F}T} \!\!\!\! 
dp \,\, \bar{\phi}(p) \, e^{+ipz/\hbar}  
\approx \phi(z)
\, ,
\label{result_1}
\end{eqnarray}
very close to the ``target'' state $\phi$, which was,
we recall, assumed to be localized in momentum space within a range
covered by the  
$\lbrack p_{0}; \, p_{0}+{\cal F}T \rbrack$  interval. 
This is the central result of our paper. 

Notice that the field amplitude (\ref{V(t)_2}) is, apart from 
an overall time-independent amplitude,  a product
of two distinct time-dependent factors. The second one 
is not specific for a particular ``target'', but only
for a given momentum window $\lbrack p_{0}; \, p_{0}+{\cal F}T \rbrack$.
This factor can be set   
once and for all for a broad class of targets. We recall that 
the purpose of this factor is to {\it compensate the quantum-mechanical
diffraction and acceleration} of the distribution being generated. 
The first factor (\ref{V(t)_22}) 
corresponds to a finite duration 
pulse whose amplitude is 
proportional to the 
target wave function $\bar{\phi}(p)$ in the momentum representation:
it allows  encoding of 
the ``target'' wave-function in the spectrum of the 
applied laser field.
%

\begin{figure}
\begin{center}
\leavevmode
\epsfxsize=0.5\textwidth
\epsffile{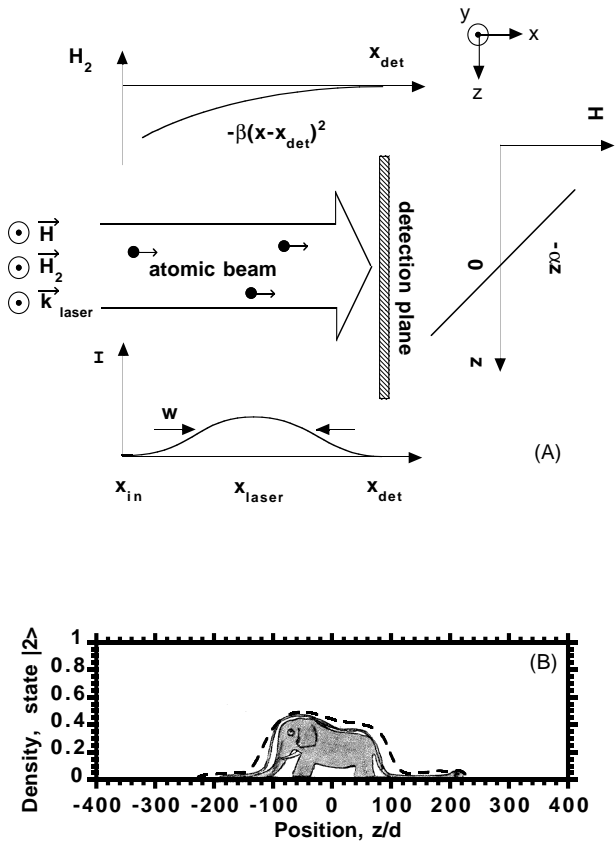}
\end{center}
\vspace{-2.5cm}
\caption
{
A possible implementation of the
``de-Broglie Wave-Front Engineering'' technique in atom lithography.
(a) Schematics of the set-up. (b) Numerical simulation of an attempt
to reproduce the ``boa swallowed an elephant'' pattern [12]
using the de-Broglie engineering technique.
Dashed line shows the 
spatial distribution of atoms in the state $|2\rangle$
in the detection plane. 
Interaction time is $T = 1. \times \tau$, Rabi frequency is 
$\tilde{V}_{0} = 1. \times \hbar/\tau$. Time width of the 
Gaussian pulse is $T_{\rm laser} = 0.3 \times \tau$. 
The pattern is represented by 
$2N +1 = 127$ pixels on a $\Delta z = 3.7 \times d$ grid. 
The distribution is a result of averaging over
20 Monte-Carlo realizations corresponding to the ``entering time''
$t_{\rm in}$ randomly distributed within an interval
$\lbrack 0; \, 2\pi/\Delta\omega \rbrack$, where
$\Delta\omega = {\cal F}\Delta z/\hbar$.
The characteristic length $d$ and
time $\tau$ of the problem are defined by the expressions
(\ref{gravitational_width}) and (\ref{gravitational_time}) respectively.
\label{f_elephant}
}
\end{figure}

Let us discuss now possible implementations of our scheme 
in the field of {\it atom lithography}.
Consider a set-up depicted in Fig.~\ref{f_elephant}a.
We assume that a beam of atoms in state $|1\rangle$ 
with a mean longitudinal velocity
$\langle v_{x,0} \rangle$ interacts with a superposition of a
Gaussian laser beam of a time-dependent amplitude 
and two magnetic fields ($z$- and $x$-dependent respectively).
The amplitude of the $x$-dependent component of the magnetic field 
$
H_{2}(x) =
- \beta
(x-x_{\rm det})^2
$
is chosen to be equal to 
$\beta =
\mu_{\rm Bohr} g\alpha^2/2M \langle v_{x,0} \rangle^2$.
The $z$-dependent component of the magnetic field is the same as 
in the previous example.

Consider first an idealized case of a monochromatic
($v_{x,0} = \langle v_{x,0} \rangle$)
ideally collimated ($v_{y,0} = v_{z,0} = 0$) atomic beam,
${\bf v}_{0}$ being the velocity of atoms in the incoming 
beam. Within the paraxial approximation and in the 
interaction picture with respect to the
$x$-dependent Zeeman shift, the  
laser field ``seen'' by a moving atom reads
\begin{eqnarray}
V(\check{t}) = \tilde{V}(\check{t} + t_{\rm in})
e^
{
- \, \pi(\check{t} - T/2)^2/T_{\rm laser}^2
  -i\int_{0}^{\check{t}}
  \omega_{2,1}(z_{s}(\check{t}^{\prime}, 0, 0)) \, d\check{t}^{\prime} 
} 
\label{V(t)_3}
\end{eqnarray}
where
$
\hbar \omega_{2,1}(z) = \hbar \omega_{2,1}(0) - {\cal F}z + R 
$, 
$R = \hbar^2 k^2/2M$ is the recoil energy, 
${\bf k}_{\rm laser} = k {\bf e}_{y}$ is the 
wave vector of the incident light,
$\check{t} = t-t_{\rm in}$ is the
time counted from the ``entering'' time $t_{\rm in}$  when the atom
enters the interaction zone $\lbrack x_{\rm in} \le x \le x_{\rm det} \rbrack$,
the left edge of the interaction zone is
conventionally defined as
$x_{\rm in} = x_{\rm det} - 2(x_{\rm det} - x_{\rm laser})$,
the interaction time is given by $T = (x_{\rm det} - x_{\rm in})/v_{x,0}$,
the ``source trajectory'' is defined by the expression (\ref{trajectory}),
and the time width of the Gaussian pulse is 
$T_{\rm laser} = \sqrt{\pi/2} w/\langle v_{x,0} \rangle$
where $w$ is the spatial waist of the Gaussian beam.

Imagine that one's goal is to generate,
by chosing a proper time dependence of the 
field amplitude $\tilde{V}(t)$, a pre-chosen ``target''
density profile $\sigma(z)$ (scaled in such a way that $\sigma(0) = 1$)
in the detector plane $x = x_{\rm det}$ (or, equivalently, at
$\check{t} = T$). The quantum state engineering recipe (\ref{V(t)_22})
does not directly apply:
according to this recipe the function
$\tilde{V}(t)$ should be a finite-duration time pulse,
and the probability for atom to join the interaction zone at the
time of the pulse would vanish for long exposure times. In other words,
the field amplitude (\ref{V(t)_3}) contains a new (as compare to 
(\ref{V(t)_2},\ref{V(t)_22})) uncontrollable 
parameter $t_{\rm in}$, randomly distributed 
within an interval $\lbrack 0; \, {\cal T} \rbrack$, 
${\cal T}$ being the ``exposure time''. We suggest now to
replace the pulse (\ref{V(t)_22})) by a {\it periodic} sequence 
of pulses.

Let us consider the following {\it ansatz} for the
field amplitude:
\begin{eqnarray}
\tilde{V}(t) =
\tilde{V}_{0} \sum_{n=-N}^{+N} \sqrt{\sigma(z_{n})} \,
                        e^{-i\omega_{n} \, t} \, ,
\label{V(t)_4}
\end{eqnarray}
where $z_{n} = \Delta z \, n$, and  $\omega_{n} = \omega_{2,1}(z_{n})$;
the space domain $\lbrack -\Delta z \, N; \, +\Delta z \, N \rbrack$
is supposed to cover the target's profile $\sigma(z)$. 
For long enough deposition times ${\cal T} \gg N/\Delta\omega$
the surface density of the
state $|2\rangle$ deposited on the detection plane will be given
by a sum of Gaussians profiles with an envelope
given by the ``target'' pattern:
\begin{eqnarray}
n(y,z) \approx  n_{0} \sum_{n=-N}^{+N} \sigma(z_{n})
          e^{ - \, \frac{(z-z_{n})^2}{2(\delta z)^2} }
\approx n_{0}  \, 
        Q(\delta z/ \Delta z) \, 
        \sigma(z)
\, ,
\nonumber
\end{eqnarray}
where the width of an individual Gaussian is given
by
$
\delta z
= \sqrt{\pi}\hbar/{\cal F}T_{\rm laser}
= \sqrt{2}\hbar \langle v_{x,0} \rangle/{\cal F}w
$, the overall density amplitude is
$n_{0} = (j_{\rm in}{\cal T})(V T_{\rm laser}/\hbar)^2$, 
$j_{\rm in}$ is the flux density in the incoming atomic beam,
and 
$
Q(\zeta) = \sum_{n^{\prime}=-\infty}^{+\infty} 
          \exp \lbrack - (n^{\prime})^2/2\zeta^2 \rbrack
$. 
Notice
that
the produced pattern does not depend on  particular values of the 
entering times 
$t_{\rm in}$ of the individual atoms, as a result of
averaging over a uniform distribution of $t_{\rm in}$
within an interval $\lbrack 0; \, {\cal T} \rbrack$. As in the previous example
the spatial resolution $\delta z$ of the atomic pattern engineering method
is limited {\it only} by the position-momentum uncertainty relation
$\delta z \sim \hbar/{\cal F} T_{\rm laser}$.

In Fig.\ref{f_elephant}b we show the result of our attempt to generate 
the ``boa swallowed an elephant'' pattern \cite{Petit_Prince},
using this method. 

For a realistic atomic beam of a finite spread in the 
longitudinal ($\delta v_{x,0}$) and transverse
($\delta v_{z,0}$ and $\delta v_{y,0}$)
velocities, the spatial resolution of the above lithographic method 
will be limited by 
$\delta_{(\delta v_{x,0})} z \sim
({\cal F} T^2/M) \, (\delta v_{x,0} / \langle v_{x,0} \rangle)$,
$\delta_{(\delta v_{z,0})} z \sim
\delta v_{z,0} T$, and 
$\delta_{(\delta v_{y,0})} z \sim
k \delta v_{y,0} / {\cal F}$, respectively. 
\begin{figure}
\begin{center}
\leavevmode
\epsfxsize=0.45\textwidth
\epsffile{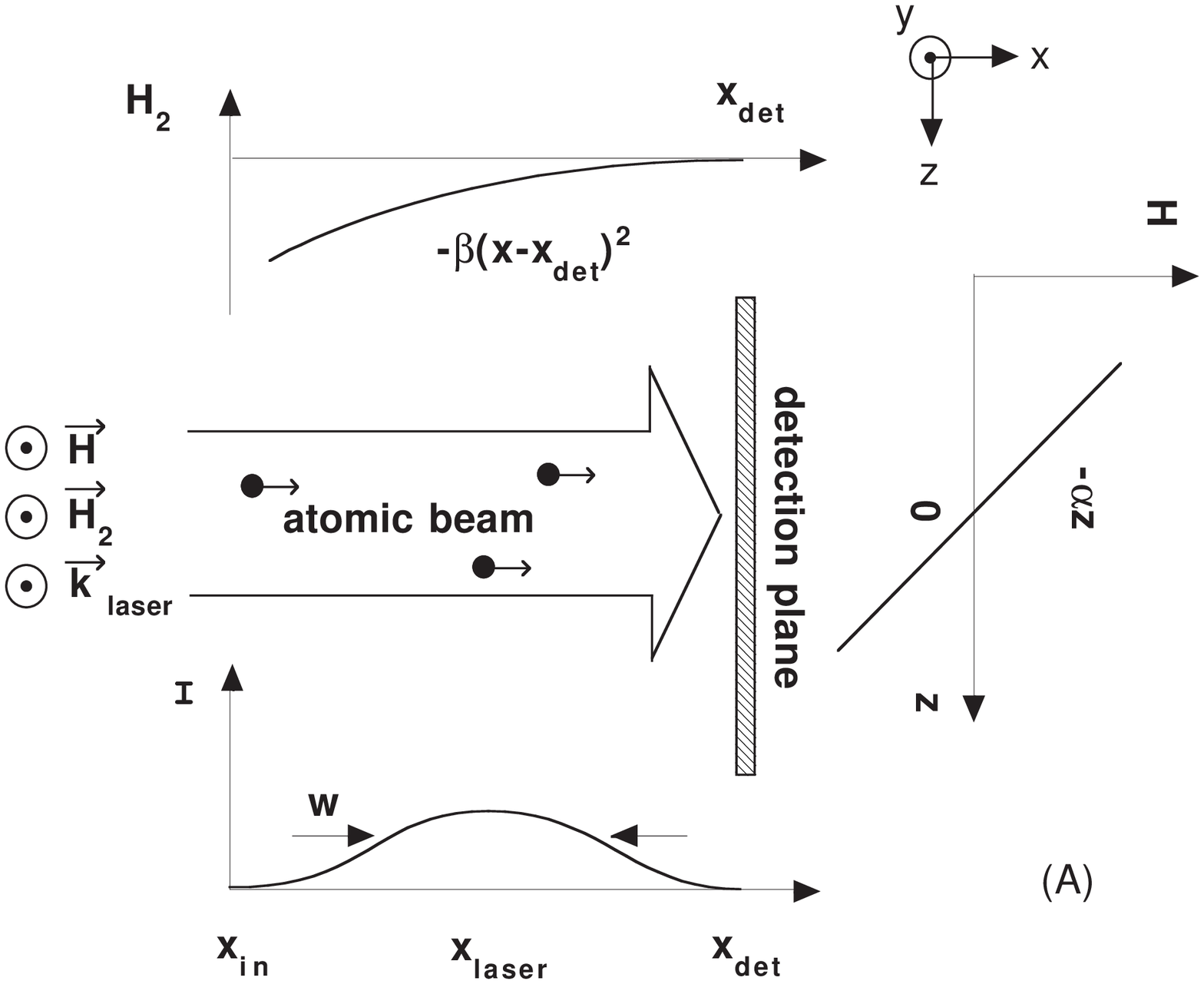}
\end{center}
%
\end{figure}
\vspace{-2.5cm}

To conclude, we have presented a method for generation of an
{\it arbitrary} motional quantum state of a free atom.
For a target state $\phi(z)$ such that its momentum 
representation $\bar{\phi}(p)$ is well localized 
within an interval $\lbrack p_{\rm min}; \, p_{\rm max} \rbrack$,
our {\it de-Broglie wave-front engineering} method allows one 
to create a state
\begin{eqnarray}
%
&&
e^{i p_{0} z/\hbar} \longrightarrow
\int_{p_{0}}^{p_{0}+{\cal F}T} \! dp \, 
\bar{\phi}(p) \, e^{ipz/\hbar}
\approx
\phi(z)
\, ,
\nonumber
\end{eqnarray}
where the initial momentum $p_{0}$ and the interaction time 
$T$ are chosen in such a way that the interval 
$
\lbrack p_{\rm min}; \, p_{\rm max} \rbrack 
$
belongs to the $\lbrack p_{0}; \, p_{0}+{\cal F}T \rbrack$ window,
${\cal F}$ being the gradient force used in course of the generation
procedure.  

We foresee that, using the classical-mechanical analogy similar
to (\ref{trajectory}), our method can be easily generalized to the case of 
trapped particles. 

We present also a modification of the method, suitable for 
lithography with atomic beams. 
The spatial resolution of the suggested lithographic method
is limited only by the position-momentum uncertainty relation,
and neither
quantum-mechanical diffraction of the pattern nor 
acceleration of atoms during the generation process affects the 
resolution.

Let us give some realistic estimates for
the spatial resolution one can achieve using the above lithographic method;
we will use Fig.~\ref{f_single} as an example.
Recall that this plot shows the
narrowest peak, which could be obtained for given values of
the field gradient ${\cal F}$ and interaction time $T$.
For the Argon mass $M = 30 {\rm amu}$ and a realistic value of the
magnetic-field-induced 
(alternatively the Stark shift in a spatially varying laser field 
induced \cite{Thomas}) gradient force
${\cal F}/\hbar = 2\pi\times 10^9 {\rm Hz}/{\rm cm} \,
(2\pi\times 10^{12} {\rm Hz}/{\rm cm})$
the natural units of length and time will be given by
$d = (\hbar^2/2 M {\cal F})^{1/3} = 108.{\rm nm} \, (10.8 {\rm nm})$ and
$\tau = (2 \hbar M/{\cal F}^2)^{1/3} = 14.7  \mu{\rm s} \, (.147 \mu{\rm s})$, 
respectively.
The HWHM of the peak shown in Fig.~\ref{f_single} will
correspond then to $63.0 {\rm nm} \, (6.3 {\rm nm})$ obtained
for an interaction time of 
$T = 4.8 \tau = 70.8 \mu{\rm s} \, (.708 \mu{\rm s)}$.
Note also that lithography techniques
sensitive to the internal state of atoms do exist and
they were  experimentally demonstrated for the case of metastable atoms
\cite{Argon}.

We would like 
to express our appreciation for many useful 
discussions with J. Eberly, K. S. Johnson, 
W. D. Phillips, B. Shore, J. H. Thywissen,
G. Zabow, and P. Zoller.
  

M.O. was supported by the National Science Foundation
grant for light force dynamics \#PHY-93-12572. 
N.D. was supported by the  NSF grant for 
Materials Research Science and Engineering Center
\#DMR-9400396.
C.H. was supported by Harvard University.
This work was also partially supported by
the NSF through
a grant for the Institute for Theoretical Atomic and Molecular
Physics at Harvard University and the Smithsonian Astrophysical 
Observatory.
%
%

%

\end{document}